\documentstyle[preprint,cite,aps]{revtex}
\begin{document}
\draft
\title{Non Relativistic Limit of a Model of  Fermions interacting
through a Chern-Simons Field}
\author{  A.J. da Silva\footnote{e-mails:{\it ajsilva@fma.if.usp.br}}}
\address{Instituto de F\'\i sica, USP\\
 C.P. 66318 - 05389-970, S\~ao\ Paulo - SP,  Brazil}
\date{December 1998}
\maketitle
\begin{abstract}
  We study the non relativistic limit of a Model of Fermions interacting
through a Chern-Simons Field, from a perspective that resembles the 
Wilson's Renormalization Group approach, instead of the more usual
approach found in most texts of Field Theory. The solution of some 
difficulties, and a new understanding of non relativistic models is given.
\end{abstract}
 
\vspace{4cm}

Models of a Chern-Simons \cite{Jackiw} field interacting with non relativistic  bosons \cite{Lozano} or fermions \cite{Hagen} have being studied in the literature both for its
interest in general understanding of field theory by itself  
as for its application to Condensed Matter Physics \cite{Zhang}. The use
of these models face in general, the difficulties of their non 
renormalizability. This fact is perhaps, the main reason for the interest,
 on the results of Bergman and Lozano \cite{Lozano},
in one loop, later extended to three loops \cite{Ruiz}.
Their model consists of a non-relativistic boson field  $\phi$  with a 
quartic self interaction and minimally interacting with a Chern-Simons 
field  $A^{\mu}$ :

\begin{eqnarray}
\label{1}
{\cal L}=& &\phi^* \biggl(i \frac{d}{dt} +e A^0\biggr)\phi - 
\frac{1}{2m}|(\vec{\nabla} -i e 
\vec{A})\phi|^2 \nonumber\\
& & -\frac{\lambda_0}{4}(\phi^* \phi)^2+\frac{\theta}{2}\,\epsilon^{\mu\nu\rho} A^{\mu} \partial^{\nu} A^{\rho}.
\end{eqnarray}
The only primitively divergent Green Function is the boson four point
function. Up to one loop, the model can be made finite by the choice of a renormalized
coupling constant $\lambda$ through the equation;
\begin{equation}
{\lambda}_0 = {\lambda}\,+\,\frac{m}{4 \pi}\biggl(\lambda^2 - \frac{4 e^2}
{m^2 \theta^2} \ln\,(\frac{\Lambda}{M})\biggr)
\end{equation}
where $\Lambda$ is an ultraviolet (UV) cut-off and $M$ an arbitrary
constant (the renormalization constant) with dimension of mass.
Their main observation is that at the critical value, ${\lambda}^*=
|\frac{2\, e}{m \theta}|$,  the one loop contribution vanishes and
no renormalization of $\lambda$ is needed. At this choice of $\lambda$
the model regains the scale invariance that it has at classical level,
and the relative wave function of the two bosons reproduces the
Aharonov-Bohm scattering amplitude \cite{Aharonov} up two the second
Bohr order.
The model of non relativistic fermions interacting with the Chern-Simons
field was also discussed in \cite{Lozano} and studied in more details in
\cite{Hagen}. In this last paper it is shown that the one loop 
scattering of two fermions with spins of the same sign (in 2+1 dimension
the spin is a pseudo-scalar) is finite in one loop, due to the contact 
interaction represented by the Pauli interaction, that is already present
in the minimal interaction of the fermions with the gauge field. As for the 
scattering of two fermions of opposite spins the Pauli interaction
does not have any role and the amplitude is divergent unless a quartic
fermionic interaction of the form  $c(\Lambda)\,\psi^*\psi
\,\phi^*\phi$  where $\psi$ and $\phi$ represent respectively fermions
with spin plus and minus $1/2$, and $c$  is a constant that depends 
logarithmically on the UV cut off $\Lambda$. 
This last fact poses a new problem. If the non relativistic model is
thought to be the low energy limit of a more fundamental model of
relativistic Dirac fermions interacting with a Chern-Simons field,
in the way that this limit is generally taken in most texts \cite{Landau},
it would come from a similar quartic interaction in the relativistic
fermions. As is well known one such interaction is non renormalizable!
We will show that this is, in fact, a false problem. No quartic non
renormalizable self interaction is needed in the ``parent'' relativistic
model if a new perspective on the non relativistic limit in field theory
is taken. Before going to the description of this new limit, lets us
briefly resume, in an example, the ``Classical Non Relativistic Limit'',
and discuss why it is not always correct.
   Let us consider, in 2+1 dimension, a 2 component Dirac fermion field 
$\Psi$, that represents a spin plus fermion and its anti-fermion, interacting
with an external electromagnetic field $A^{\mu}$, as described by the Lagrangian density (the gamma matrices are $\gamma^0\doteq 
\sigma^0$ ,$\gamma^1\doteq i\sigma^1$ and $\gamma^2 \doteq i\sigma^2$ where
$\sigma^\mu$ are the Pauli matrices)
\begin{eqnarray}
\label{2}
{\cal{L}}_{rel} = \bar{\Psi}\,\left\{\, \gamma^\mu 
\left( i \frac{\partial}{\partial x^\mu} 
+ e A^\mu\, \right)\,- m\,\right\}\,\Psi.
\end{eqnarray}
  The corresponding equation of motion is:
\begin{eqnarray}
\label{3}
\left(\,i \frac{d}{dt}+ e A^0\,\right)\,\Psi = \left\{\gamma^0 \vec{\gamma}.
\left(\,-i \vec{\nabla} -e \vec{A}\,\right)+\gamma^0 m \right\}\,\Psi.
\end{eqnarray}
   Let us now consider a positive energy solution $\Psi$ of this equation.To
make contact with the non relativistic description,in which, the rest energy 
$m$ of the particles is not included in the solution,let us make in the 
above equation of motion ,the substitution
\begin{equation}
\label{4}
\Psi\,=\,\frac{ e^{-imt}}{\sqrt{2 m}} \left(\begin{array}{c}\psi\\ 
\chi\end{array}\right).
\end{equation}   
   The result is:
\begin{eqnarray}
\label{5}
\left( i \frac{d}{dt}+e A^0 \right)\psi&=&i\Pi_{-}\chi\\
\left( i \frac{d}{dt}+e A^0+2m \right)\chi&=&-i\Pi_{+}\psi
\end{eqnarray}
where  $\Pi_{\pm}\,=\,\Pi^1\,{\pm}\,\Pi^2$ , and  $\Pi^i=-i\frac{d}{dx^i}-eA^i$.  If we make the assumptions that: $e|A^0|<<m$ and that the momentum space 
components of $\psi$ and $\chi$ are non null only for low momenta and energies,
$(|\vec{p}|,E)<<m$ , then the second equation can be approximately solved for
$\chi$ , and inserted in the first one, giving: 
\begin{equation}
\label{6}
\left(i \frac{d}{dt}+e A^0 \right)\psi=\frac{1}{2m}\left(-i\vec{\nabla}-e\vec{A}\right)^2 + \frac{e}{2m} B\, \psi,
\end{equation}
where $B\,=\,\vec{\nabla}\wedge\vec{A}$ is the magnetic field. The one component spinor $\psi$ represent a fermion with spin plus. The last term  is the Pauli magnetic moment--magnetic field interaction term of . The Lagrangian
density corresponding to this equation  of motion is the so called Pauli 
Schr\" odinger ( PS ) propagator of non relativistic fermions in an electromagnetic field:
\begin{equation}
\label{7} 
{\cal{L}}_{nonrel}^{class}=\psi^* \left(i\frac{d}{dt}+eA^0\right)\psi-
\frac{1}{2m}|\left(-i\vec{\nabla}-e\vec{A}\right)\psi|^2 +\frac{e}{2m} B \psi^*
\psi
\end{equation}
   The essential facts behind the above non relativistic limit are the
assumptions on the strength of $A^{\mu}$  and the momentum space support of 
the field $\Psi$ (the second assumption is not meaningfull without the first, since  a low energy initial state of $\Psi$ can be driven to a relativistic state by the action of a strong $A^{\mu}$ field ). Suppose now that $A^{\mu}$ is not an external,
controllable field, but is a dynamical field with dynamics given by a Maxwell
or Chern-Simons term (that must be thought as added to the Lagrangian (\ref{2})). Let us consider in this theory, the scattering of two low energy fermions, 
their energy and momenta given by 
$\left(p^0= m+\frac{|\vec{p}|^2}{2m}, {\pm}\vec{p}\right)$ with $|\vec{p}|
<<m$ . On the right side of Figure 1 we draw a possible one loop contribution
(among others) to this process. The corresponding amplitude is given by a
Feynman integral in the loop momentum $k^{\mu}$:
\begin{equation}
\label{8} 
{\cal{A}}_{lowenergy}^{relativ}\,=\,\cdots\int^\infty d^3 k\left(\,\cdots\,\frac{i}{(k^0+p^0)^2-
(\vec{k}+\vec{p})^2 -m^2}\, \right)\,\cdots  
\end{equation}
   The main observation on this equation is that, even if the process we are
treating is a low energy one, the amplitude receives contributions of high
energy intermediate states, represented by propagators whose dynamics is
essentially relativistic, and so, not coming from the Feynman rules of the non
relativistic Lagrangian (\ref{7}). The bigger or lower suppression of the
contribution of these high energy states depend on the dynamics of the
exchanged $A^{\mu}$ field. As we will explicitly see in one example below,
for the Chern-Simons field, they effectively give a contribution that can not
be understood as coming from the non relativistic Lagrangian (\ref{7}).    
   What about the description of this same process starting from the non
relativistic theory given by (\ref{7})? The amplitude for the same process of
figure 1 is of the form 
\begin{equation}
\label{10} 
{\cal{A}}^{class}_{nonrel}\,=\,\cdots\int^\infty d^3 k\left(\,\cdots\,\frac{i}{(k^0+p^0)-(\vec{k}+\vec{p})^2/2m }\, \right)\,\cdots  
\end{equation}
where now $p^0=|\vec{p}|^2/2m$. Here also, the integral extends up to
infinity momenta. High energy intermediate states also contribute to the
amplitude, even with a bigger weight than to ${\cal{A}}_{lowenergy}^{relativ}$, as seen from the worse UV behavior of the Pauli-Schr\"odinger(PS) propagator.
It must yet be observed, that from the view point of the Special Relativity,
the PS propagator misses to represent the propagation of high energy 
intermediate states. Due to these facts, some authors in Field Theory
(\cite{Lepage}) take
 the view of Non Relativistic Field Theory
as a cut off theory. This means that instead of considering ,in a wrong way,
the contribution of the high energy intermediate states ,they prefer to
decouple them from the theory by limiting the integration in the Feynman
integrals up to a maximal energy-momentum,
compatible with the newtonian description provided by the PS propagator.
This is also a view  taken by some authors in optics (\cite{Cohen}). There, the
typical energy involved in the scattering processes are of the order
of the ionization energy of the atoms, $\alpha^2\, m$, where $\alpha$ is the fine structure constant and $m$ is the electron mass. The cut off assumed is $\Lambda\,=\,\alpha\, m$,
the inverse of the Bohr radius of the atom, much bigger than the typical 
energies involved in the scattering processes, and much smaller than the
rest energy, $m$ of the electron.

    We will take a slight variation of these ideas, suited for 
understanding the results on non relativistic models  with a CS field  in the Coulomb Gauge, as treated in the literature \cite{Lozano,Hagen}. We will consider a non relativistic  cutoff,  only in the spatial momentum $\vec{k}$, of the
Feynman integrals; that is, we will calculate the Feynman integrals,firstly 
 freely integrating the $k^0$ momentum component up to infinity, and
then restricting the integration in $|\vec{k}|$ to the region $(0\,,\,
\Lambda)$ with $\Lambda$ chosen so that $|\vec{p}|\,<<\,\Lambda\,<<\,m$,
 where $\vec{p}$ is a characteristic momentum of the low energy process 
in which  of interest, and $m$ is the mass of the fermion field.  
This choice has the additional technical advantage of not breaking the
locality in the time direction. The way, we are proposing, that  
these cut off non relativistic models are  related to originally 
relativistic ones, is akin to the ideas of the Renormalization 
Group of Wilson \cite{Wilson}.
We will first outline the main ideas in a toy model.  
  
Lets us consider a relativistic field theory in 1 space-time dimension, with
dynamics given by a Lagrangian ${\cal{L}}^{rel}(\Phi)$. Its functional
generator is given by
\begin{equation}
\label{11}
{\cal{Z}}(j)=\int {\cal{D}}\Phi(p)\, \exp \left(i\int dp\left({\cal{L}}^{rel}(\Phi)+j\Phi \right)\right)
\end{equation}
where $j$ is an external source, and  ${\cal{D}}\Phi(p)\doteq \prod_0^{\infty}
d\Phi(p)$. Suppose that we are only interested in describing ``non 
relativistic''processes 
involving external particles with momenta $p$ smaller than a certain value
$\Lambda<<m$, where $m$ is the mass of the field $\Phi$. This limitation can be implemented in the functional generator by choosing the external source to be non null only for the momentum region $(0\,,\,\Lambda)$. The $\Phi$ field can be separated in  $\Phi(p)\,=\,\phi(p)\,+\,\bar{\phi}(p)$ where $\phi$ 
has support in $(0,\Lambda)$  and $\bar{\phi}$ in $(\Lambda,\infty)$. 
The integration
measure goes in ${\cal{D}}\Phi\,=\,{\cal{D}}\phi\,{\cal{D}}\bar{\phi}$,  the
Lagrangian separates in ${\cal{L}}^{rel}(\Phi)\,=\,{\cal{L}}^{rel}(\phi\,+\,\bar{\phi})\,=\,{\cal{L}}^{rel}(\phi)\,+\,\nabla{\cal{L}}(\phi\,,\,\bar{\phi})$
and $j\,\Phi$ gets reduced to $j\,\phi$. The functional generator becomes
\begin{equation}
\label{12}
{\cal{Z}}(j)\,=\,\int{\cal{D}}\phi\,\exp \,i\int \left({\cal{L}}^{rel}(\phi)\,+\,j\,\phi\right)\,\int {\cal{D}}\bar{\phi}\,\exp\, i\int \left(\nabla{\cal{L}}(\phi,\bar{\phi}\right)
\end{equation}
and can be written in the form
\begin{equation}
\label{13}
{\cal{Z}}(j)\,=\,\int{\cal{D}}\phi\,\exp \,i\int \left({\cal{L}}^{effect}(\phi,\Lambda)\,+\,j\,\phi\right)
\end{equation}
where  ${\cal{L}}^{effect}(\phi\,,\,\Lambda)\,=\,{\cal{L}}^{rel}(\phi)\,+\,
\delta{\cal{L}}(\phi\,,\,\Lambda)$   and
\begin{equation}
\label{14}
\int  \delta{\cal{L}}(\phi\,,\,\Lambda)\,=\,-i\ln \left(\int{\cal{D}}\bar{\phi}\, \exp\,i\,\int \nabla{\cal{L}}(\phi\,,\,\bar{\phi})\right)
\end{equation}
The effects of the high momenta modes $\bar{\phi}$  are incorporated in the 
effective dynamic of the low energy ones, through the additional term
$\delta{\cal{L}}(\phi\,,\,\Lambda)$. It is the effective Lagrangian, 
${\cal{L}}^{effect}$  , in which
the only remaining free momemta modes are the non relativistic ones, and not
the original ${\cal{L}}^{rel}$ , that will give, through the approximations 
called Classical Non Relativistic Limit (exemplified above), the same results
to low energy processes, as if calculated from the original relativistic 
model.

The integration in $\bar{\phi}$ in (\ref{14}) can in general be done by 
expanding the exponential in a series of powers of 
$\int \nabla{\cal{L}}(\phi\,,\,\bar{\phi})$. The result will be a series of 
Feynman graphs with the propagator of $\bar{\phi}$ in the internal lines 
and the field $\phi$ in the external legs. This means that the integration in the loop momenta is restricted to the interval $(\,\Lambda\,,\,\infty\,)$. 
The result is in general $\Lambda$ 
dependent (as we will see the result of Bergman and Lozano is an exception)
resulting in an  Effective Lagrangian ${\cal{L}}^{effect}$  
that is dependent on $\Lambda$.

Let us now return to the models that we want to treat in 2+1 dimensions. We will
start with the relativistic Lagrangian
\begin{eqnarray}
\label{15}
{\cal{L}}^{relat}&=&\bar{\Psi}\,\left(\,\gamma^{\mu}\,\left(\,i\,
\frac{\partial}{\partial x^{\mu}}\,+\,e\,A_{\mu}\,\right)\,-\,m\,\right)
\Psi\nonumber\\
&+&\bar{\Phi}\,\left(\,\gamma^{\mu}\,\left(\,i\,
\frac{\partial}{\partial x^{\mu}}\,+\,e\,A_{\mu}\,\right)\,+\,m\,\right)
\Phi\,+\,\frac{\theta}{2} \epsilon_{\mu\nu\rho}\,A^{\mu}\,\partial^{\nu}\,
A^{\rho}
\end{eqnarray}
where $\Psi\,\,(\Phi)$  is a 2 component Dirac field representing a fermion
and anti fermion of spin $plus\,\, (minus)$.
In the Coulomb Gauge, the CS propagator is (indices $\mu,\nu,\dots $ runs
from 0 to 2 and indices $i,j,\dots$ runs over 1 and 2)
\begin{eqnarray}
\label{16} 
\Delta_{\mu\nu}&\doteq&<TA_{\mu}(p)A_{\nu}(-p)>\,=\,\frac{1}{\theta}
\,\epsilon_{\mu\nu i}\,\frac{k^i}{\vec{k}^2}\\
\end{eqnarray}
and will be represented by a wavy line. The Dirac propagators of the 
relativistic fermions will be represented by double straight lines.
Through the same steps that led
(2) to (7) we get the Classical Non Relativistic limit of this model
\begin{eqnarray}
\label{17} 
{\cal{L}}^{class}_{nonrel}&=&\psi^* \left(i\frac{d}{dt}+eA^0\right)\psi-
\frac{1}{2m}|\left(-i\vec{\nabla}-e\vec{A}\right)\psi|^2 +\frac{e}{2m} B \psi^*
\psi\nonumber\\
&+&\phi^* \left(i\frac{d}{dt}+eA^0\right)\phi-
\frac{1}{2m}|\left(-i\vec{\nabla}-e\vec{A}\right)\phi|^2 -\frac{e}{2m} B \phi^*
\phi\nonumber\\
&+&\frac{\theta}{2} \epsilon_{\mu\nu\rho}\,A^{\mu}\,\partial^{\nu}\,
A^{\rho},
\end{eqnarray}
where $\psi\,\,(\phi)$ are anti commuting one-component fields representing
a spin $ plus (minus)$ fermion. The fermionic PS propagator will be represented by a single straight line. This model has several 
different vertices : $F^*FA^0$, $F^*F\vec{A}$, $F^*FA^{\mu}A_{\mu}$
and $F^*FB$, where F stands for $\phi$ or $\psi$. The last vertex (Pauli)  gives a local interaction between two
fermions, mediated by the the propagator
\begin{eqnarray}
\label{18}
\Delta_B&\doteq&<TB(k)A_0(-k)>=\frac{i}{\theta}
\end{eqnarray}
that we will represent by an dashed straight line. The fermionic 
PS Propagator will be represented by a single straight line.  

We will leave the result above for future use, and 
return to the construction of the Effective Non Relativistic Model. This will
be done by calculating different low energy processes in the Relativistic
Theory and identifying the contributions that come from the low energy intermediate states (and are the same that come from 
the Classical Non Relativistic Model with a cut off $\Lambda$) and the contributions that come from high energy 
intermediate states, and  that must be incorporated in the Effective Non Relativistic Model, through new terms in the Lagrangian.  We will restrict the calculations
to the one loop order. The sum of Feynman graphs, written as a loop integral 
can be separated as
\begin{equation}
\label{19} 
{\cal{A}}_{lowenergy}^{relativ}\,=\,\int^\infty d^3 k I\left(k^0,\vec{k},w(p),\vec{p} \right)= \int^{\Lambda}_{0}d^2k \int^{\infty}_{-\infty}dk^0 I
+\int_ {\Lambda}^{\infty}d^2k \int^{\infty}_{-\infty}dk^0 I  
\end{equation}
In the low momenta part, both $|\vec{p}|$ and $|\vec{k}|$ are smaller than 
$\Lambda<<m$, and we can safely make the approximation $w(\vec{q})=m+\frac
{\vec{q}^2}{2m}$, for both p and k. The propagators and vertices collapse
in the correspondent ones, got from the Lagrangian (\ref{15}).
In the high 
intermediate energy part this approximation can be taken for w(p)
but not for w(k). As $|\vec{p}|<<\Lambda$ and the integral  is for 
$|\vec{k}|>\Lambda$, the result, $H(p,\Lambda)$, is analytic in p and can  
be expanded in a series in p. Every term of this expansion is a
contribution to the process, that can be represented by a (new) local 
term in the Lagrangian of the Effective Non Relativistic Model.
The three processes that require renormalization are the Vacuum Polarization
Tensor (Figure 2) the Fermion Self energy (Figure 3) and the Vertex 
Correction (Figure 4). The calculation, of these quantities in covariant gauges are presented in many papers in the literature (\cite{Kogan}).
In the Coulomb Gauge it was obtained in \cite{Hagen3,Fleck}.
The results, separating the contributions of the low (first bracket)
and of the high (second bracket) intermediate momenta contributions are
respectively
\vspace{1cm}
\begin{eqnarray}
\label{20}
& &\Pi_{\mu\nu}^{lowenergy}=\Biggl[\,Zero+O(1/m^2)\,\Biggr]+\Biggl[-i\frac{e^2}{6\pi m}
(p^2 g_{\mu\nu} - p_{\mu}p_{\nu})+O(1/m^2)\Biggr]\\
\,\,\,\nonumber \\
& &\Sigma^{lowenergy}_{\psi or \phi}=\Biggl[\,Zero+O(1/m^2)\,\Biggr]+\Biggl[\,\,i\,\frac{e^2}{4\pi \theta}\,(\,{\pm}\vec{\gamma} \cdot \vec p\,-\,\frac{{\vec p}^2}{m})\,+O\,(1/m^2\,\,)\Biggr]\\ \vspace{1cm}
\,\,\, \nonumber \\
& &e\,A^{\mu}_{external}\,\bar{u}(p')_{\psi or \phi}\Gamma^{lowenergy}_{\mu}
(p'-p)\ u(p)_{\psi or \phi}\nonumber\\ \vspace{1cm}
& &\,\,\,\,\,\,\,\,\,\,\,\,\,\,\,\,\,\,\,\,\,\,\,\,\,\,=\Biggl[\, Zero +O(1/m^2)\,\Biggr]+\frac{e}{2m\,}\Biggl[\,\frac{e^2}{2\pi \theta}\epsilon^{ij}\frac{(p'-p)^j}{2m}
A^i_{external}+O(1/m^2)\Biggr] 
\end{eqnarray} \vspace{1cm}

As indicated in these formulas, all the contributions to these functions come from the high 
momenta intermediate states.
In fact it is well known that these same functions are zero when 
calculated in the classical non relativistic model(\cite{Lozano}). 
As consequence the
whole contribution to these functions, come only from the high momenta
intermediate states and are independent of the cut off $\Lambda$. The effects
of these terms in correcting the low energy dynamics of the fermions and
the CS field are simulated by adding to the Lagrangian (\ref{17}) the terms
\begin{eqnarray}
\label{21}
\delta{\cal{L}}=-\frac{1}{4}\frac{e^2}{6 \pi m} F_{\mu\nu}F^{\mu\nu}+
\frac{e}{2 m}\frac{e^2}{2 \pi \theta} B \psi^* \psi+
\frac{e}{2 m}\frac{e^2}{2 \pi \theta} B \phi^* \phi.
\end{eqnarray}
From (\ref{21}) and (\ref{17}) we see that the CS field becomes a
dynamical propagating field, the so called Maxwell-Chern-Simons field (\cite{Jackiw}). We can also see that the magnetic momenta
of the spin $plus$ and $minus$ fermions are corrected to (\cite{Fleck})
\begin{eqnarray}
\label{22}
\mu_{\psi or \phi}=\frac{e}{2 m}\left({\pm}1 +\frac{e^2}{2 \pi \theta}\right)
\end{eqnarray}
(these results where obtained previously in the literature in covariant
gauges(\cite{Kogan})).

Let us now look at the elastic scattering of two low energy  
fermions. For simplicity we will work in the Center of Momentum Reference
Frame in which the incoming fermions have energy and momenta:
$(m+\frac{\vec{p}^2}{2 m},\vec{p}\,)$ and $(m+\frac{\vec{p}^2}{2 m},-\vec{p\,)}$ 
and the outgoing fermions have 
$(m+\frac{\vec{p'}^2}{2 m},\vec{p'}\,)$ and $(m+\frac{\vec{p'}}{2 m},-\vec{p'}\,)$ with $|\vec{p}|=|\vec{p'}|<<\Lambda$.
The amplitude is a function of $|\vec{p}|$ and the angle between $\vec{p}$
and $\vec{p'}$. We prefer to write it in terms of
$\vec{p}$ and the two vectors $\vec{s}\doteq\vec{p}+\vec{p'}$ and $\vec{q}
\doteq\vec{p'}-\vec{p}$.
In Figure 5 are shown the non null graphs contributing to this process.

For the scattering of one fermion of spin $plus$ and other of spin $minus$, 
the contributions of these graphs are listed below, separated
in two rows. In the first 
 are the contributions of the 
low intermediate momenta states, ${\cal{A}}(0,\Lambda)$, and in the second 
row, the local (independent of p) contributions of the high momenta 
intermediate states,  ${\cal{A}}(\Lambda,\infty)$.
\vspace{1cm}
\begin{equation}
\label{23}
\begin{array}{lcccc} \vspace{1cm}
{\cal{A}}^{\,++\,\,\,relat}_{lowenergy}\,=&{\cal{A}}^{\,++\,rel}_{lowene}(0,\Lambda)&+&{\cal{A}}^{\,++\,rel}_{lowene}(\Lambda,\infty)\\ \vspace{1cm}
Graph\,\, 5a = \Biggl[&i\frac{e^2}{m \theta}\frac{\vec{s}\wedge\vec{q}}
{\vec{q}^2}& \Biggr]\,+\,\Biggl[&0& \Biggr]\\ \vspace{1cm}
Graph\,\, 5b = \Biggl[&0& \Biggr]\,+\,\Biggl[&\frac{e^4}{6 \pi m \theta^2}
& \Biggr]\\ \vspace{1cm}
Graphs\, 5c = \Biggl[&0& \Biggr]\,+\,\Biggl[&\frac{e^4}{2 \pi m \theta^2}
& \Biggr]\\ \vspace{1cm}
Graph\,\, 5d = \Biggl[&\frac{e^4}{4 \pi m \theta^2}\ln\frac{-\vec{q}^2}{\vec{p}^2}
& \Biggr]\,+\,\Biggl[&0& \Biggr]\\ \vspace{1cm}\
Graph\,\, 5e = \Biggl[&\frac{e^4}{4 \pi m \theta^2}\ln\frac{\Lambda^2}{\vec{q}^2}
& \Biggr]\,+\,\Biggl[&\frac{e^4}{4 \pi m \theta^2}\ln\frac{4 m^2}{\Lambda^2}
& \Biggr]\\                                                       
\end{array}
\end{equation}

Some observations are in order:
1. The ${\cal{A}}(0,\Lambda)$ parts
of each graph ( of the Relativistic 
Model ) are the same as calculated from the Classical Non Relativistic 
Model (\ref{17}) with a cut off $\Lambda$, through the graphs drawn on 
Figure 5, at the right of the corresponding relativistic ones. 
 2. The ${\cal{A}}(0,\Lambda)$ part of each graph can depend on the non
relativistic cut off $\Lambda$ (see 5e) but the whole graph 
is independent of $\Lambda$ ,as can be seen by adding ,
for each graph, the terms 
of the first and the second row.
3. Had we interpreted $\Lambda$ as an UV cut off in the usual way, i.e.  
$\Lambda\longrightarrow\infty$, and  ${\cal{A}}(0,\Lambda)$
would be a divergent amplitude. 
4. The ${\cal{A}}(\Lambda,\infty)$ non null contributions of the graphs 
5b and 5c could also be get by calculating 5a, starting
from the already corrected Effective Lagrangian given by (\ref{17})
plus (\ref{25}). 
5. The non null  ${\cal{A}}(\Lambda,\infty)$ part of diagram 5e
instead, is a new term that must be incorporated in the Effective
Lagrangian as a local quartic interaction of the form 
$\psi^*\psi\,\phi^*\phi$. It must be stressed that this term comes from
the integration over the high momenta intermediate states of the Renormalizable
Relativistic Model; no quartic term of the form $\Psi^*\Psi\,\Phi^*\Phi$
is needed in the ``parent''Relativistic Model to generate this quartic term  in
the Effective Non Relativistic Lagrangian. 
The Effective Non Relativistic Model incorporating all these terms can be written
\begin{eqnarray}
\label{24} 
{\cal{L}}^{effect}_{nonrel}&=&\psi^* \left(i\frac{d}{dt}+eA^0\right)\psi-
\frac{1}{2m}|\left(-i\vec{\nabla}-e\vec{A}\right)\psi|^2 +\frac{e}{2m}(1+\frac{e^2}{2 \pi \theta}) B \psi^*
\psi\nonumber\\
&+& \phi^* \left(i\frac{d}{dt}+eA^0\right)\phi-
\frac{1}{2m}|\left(-i\vec{\nabla}-e\vec{A}\right)\phi|^2 +\frac{e}{2m}
(-1+\frac{e^2}{2 \pi \theta}) B \phi^*
\phi\nonumber\\
&+&\frac{\theta}{2} \epsilon_{\mu\nu\rho}\,A^{\mu}\,\partial^{\nu}\,
A^{\rho}-\frac{1}{4}\frac{e^2}{6 \pi m} F_{\mu\nu}F^{\mu\nu}\nonumber\\
&+&\left(\frac{e^4}{4 \pi m \theta}\ln\frac{4m^2}{\Lambda^2}\right)
\psi^*\psi\,\phi^*\phi.
\end{eqnarray}
The calculation of the magnetic moment of the fermions, the propagator of 
the (Maxwell) Chern-Simons, and of the low energy scattering of two fermions,
in this theory, using a cut off $\Lambda$ (up to one loop), give the same
results as the calculation of the same quantities starting from the 
Relativistic Model (\ref{15}). For example, the amplitude of scattering
of one spin $plus$ and one spin $minus$ fermion ( the sum of the two rows
in equation (\ref{23})) gives (\cite{GS})
\begin{equation}
\label{25}
{\cal{A}}^{\,++\,\,eff}_{nonrel}\,\doteq\,{\cal{A}}^{\,++\,rel}_{lowene}\,=\,
i\frac{e^2}{m \theta}\frac{\vec{s}\wedge\vec{q}}
{\vec{q}^2}+\frac{2 e^4}{3 \pi m \theta^2}
+\frac{e^4}{4 \pi m \theta^2}\ln\Biggl(\frac{-4 m^2}{\vec{p}^2}\Biggr)
\end{equation}
The calculation starting from the classical Non Relativistic Model,
(the sum of terms in the first row in equation (\ref{23})) would instead, 
give the ``divergent'' result (\cite{Hagen})
\begin{equation}
\label{26}
{\cal{A}}^{\,++\,class}_{nonrelat}\,=\,
i\frac{e^2}{m \theta}\frac{\vec{s}\wedge\vec{q}}
{\vec{q}^2}\,+\,\frac{e^4}{4 \pi m \theta^2}\ln\Biggl(\frac{-\Lambda^2}{\vec{p}^2}\Biggr)
\end{equation}

These results exemplify our main point: taking the non relativistic
limit in the Lagrangian and equations of motion (Classical Non 
Relativistic Limit) and then calculating a process gives in general,
a result different than, first calculating the same process 
in the relativistic theory and later taking the non relativistic 
limit  of the result.

To finish this talk I will turn to the problem that motivated this
study: the finite result for ${\cal{A}}^{class}_{nonrel}$ got
in (\cite{Lozano}) for the scattering of two bosons and its
extension (\cite{Hagen}) to the scattering of $two$  
spin $plus$ fermions (we will think that the two fermions are not 
identical and we don't need to anti symmetrize the amplitude with respect
to the outgoing particles). The non null graphs contributing to this
process are the same of figure 5.
The result is
\vspace{1cm}
\begin{equation}
\label{27}
\begin{array}{lcccc} \vspace{1cm}
{\cal{A}}^{\,-+\,\,\,relat}_{lowenergy}\,=&{\cal{A}}^{\,-+\,rel}_{lowene}(0,\Lambda)&+&{\cal{A}}^{\,-+\,rel}_{lowene}(\Lambda,\infty)\\ \vspace{1cm}
Graph\,\, 5a = \Biggl[&\frac{e^2}{m \theta}\bigl(\,1\,+\,i\frac{\vec{s}\wedge\vec{q}}
{\vec{q}^2}& \Biggr]\,+\,\Biggl[&0& \Biggr]\\ \vspace{1cm}
Graph\,\, 5b = \Biggl[&0& \Biggr]\,+\,\Biggl[&\frac{e^4}{6 \pi m \theta^2}
& \Biggr]\\ \vspace{1cm}
Graphs\, 5c = \Biggl[&0& \Biggr]\,+\,\Biggl[&\frac{e^4}{2 \pi m \theta^2}
& \Biggr]\\ \vspace{1cm}
Graph\,\, 5d = \Biggl[&\frac{e^4}{4 \pi m \theta^2}\ln\frac{\vec{q}^2}{\Lambda^2}
& \Biggr]\,+\,\Biggl[&\frac{e^4}{4 \pi m \theta}\ln\frac{4 m^2}{\Lambda^2}& \Biggr]\\ \vspace{1cm}\
Graph\,\, 5e = \Biggl[&-\,\frac{e^4}{4 \pi m \theta^2}\ln\frac{\Lambda^2}{\vec{q}^2}
& \Biggr]\,+\,\Biggl[&\frac{e^4}{4 \pi m \theta^2}\Biggl(\ln\frac{4 m^2}{\Lambda^2}\,-\,2 \Biggr)
& \Biggr]\\                                                       
\end{array}
\end{equation}
The differences of these results to the ones in (\ref{23}) come from
the Pauli interaction of the magnetic field of each fermion with the
magnetic moment of the other fermion. The effects of these interactions
cancel in the scattering of a spin $plus$ and a spin $minus$ fermion
and add in the case of $two$ spin $plus$ fermions. The results for
${\cal{A}}^{effect}$ and  ${\cal{A}}^{class}_{nonrel}$ are  now
\begin{eqnarray}
\label{28}
{\cal{A}}^{\,-+\,eff}_{nonrel}\,\doteq\,{\cal{A}}^{\,-+\,rel}_{lowene}\,&=&\,
\frac{e^2}{m \theta}\Bigg(\,1\,+\,i\frac{\vec{s}\wedge\vec{q}}
{\vec{q}^2}\Biggr)\,+\,\frac{ e^4}{6 \pi m \theta^2} \\
\,\,\,\nonumber \\
{\cal{A}}^{\,-+\,class}_{nonrelat}\,&=&\,
\frac{e^2}{m \theta}\Biggl(\,1\,+\,i\frac{\vec{s}\wedge\vec{q}}         
{\vec{q}^2}\,\Biggr)
\end{eqnarray}
The unexpected fact that this last result is finite, independent of $\Lambda$,
is in the literature (\cite{Lozano}) related to the preservation 
at quantum level, of the scale invariance that the classical
non relativistic model presents. In the model of bosons interacting with the CS field this only happens for the special value of the quartic self interaction
discussed in the introduction. For fermions the same fact is provided by the Pauli interaction which already appear in the minimal interaction with the CS field; no fine tuning of coupling constants is needed. We here showed another aspect of this independence of $\Lambda$. Unusually,not only 
${\cal{A}}^{relat}_{lowen}$ is independent of $\Lambda$: their high and low momenta intermediate energy contributions  are separately 
independents of $\Lambda$. So the difference of the amplitudes got from the Classical or the Effective Non Relativistic Models is only a constant independent of $\Lambda$. If the fermions are identical we must anti symmetrize
the amplitudes (\ref{28}) and (29) in the outgoing particles. In this case no difference at all appears in the final result. The amplitude got from both 
(\ref{28}) and (33) is : 
$i\,\frac{2\,e^2}{m \theta}\frac{\vec{s}\wedge\vec{q}}
{\vec{q}^2}$ , and gives the Aharonov Bohm scattering amplitude for two identical fermions.

\,\,\,\,\,\,\,\,\,\,\,\,\,\,\,AKNOWLEDGEMENTS
This work was partially supported by CNPq (Conselho Nacional de Desenvolvimento
Cientifico e Tecnologico) do Brasil. The author thanks
 M. Gomes for critically reading the manuscript.

\newpage

\begin{center}

{\bf Figure captions}
\bigskip
\bigskip

\begin{itemize}

\item Figure 1. Vacuum polarization. The double line represent Dirac
fermion propagators, and the wavy line the CS propagator.

\bigskip
\bigskip

\item Figure 2. fermion self energy.

\bigskip
\bigskip

\item Figure 3. Contributions ( in the Coulomb Gauge) to the scattering
of a fermion by an external field $A^\mu_{ext}$.
The action of the external field is represented by a cross.

\bigskip
\bigskip
\item Figure 4. Example of a one loop graph contributing to the scattering
of two fermions.

\bigskip
\bigskip
\item Figure 5. Non null graphs contributing to the scattering
of two Dirac fermions. On the right of the diagrams are represented
the correspondent graphs  in the classical non relativistic model.

\end{itemize}

\end{center}

\end{document}